# Detecting substeps in the rotary motors of $F_o F_1$-ATP synthase by Hidden Markov Models

N. Zarrabi[a], M. G. Düser[a], R. Reuter[a], S. D. Dunn[b], J. Wrachtrup[a], M. Börsch*[a]

[a]3rd Institute of Physics, University of Stuttgart, Pfaffenwaldring 57, 70550 Stuttgart, Germany;
[b]Department of Biochemistry, University of Western Ontario, London, Ontario, Canada N6A 5C1

## ABSTRACT

$F_o F_1$-ATP synthase is the enzyme that provides the 'chemical energy currency' adenosine triphosphate, ATP, for living cells. The formation of ATP is accomplished by a stepwise internal rotation of subunits within the enzyme. We monitor subunit rotation by a single-molecule fluorescence resonance energy transfer (FRET) approach using two fluorophores specifically attached to the enzyme. To identify the stepsize of rotary movements by the motors of ATP synthase we simulated the confocal single-molecule FRET data of freely diffusing enzymes and developed a step finder algorithm based on 'Hidden Markov Models' (HMM). The HMM is able to find the proximity factors, P, for a three-level system and for a five-level system, and to unravel the dwell times of the simulated rotary movements. To identify the number of hidden states in the system, a likelihood parameter is calculated for the series of one-state to eight-state HMMs applied to each set of simulated data. Thereby, the basic prerequisites for the experimental single-molecule FRET data are defined that allow for discrimination between a 120° stepping mode or a 36° substep rotation mode for the proton-driven $F_o$ motor of ATP synthase.

**Keywords:** Rotary motor, $F_o F_1$-ATP synthase, FRET, single-molecule, Hidden Markov Model

## 1. INTRODUCTION

The biological function of $F_o F_1$-ATP synthase is the catalysis of ATP formation from ADP and phosphate. Therefore, the enzyme utilizes the difference of the electrochemical potential of protons (or sodium ions in some organisms) across a lipid membrane, i.e. a pH difference plus an electric potential generated by ion concentration differences. $F_o F_1$-ATP synthases are embedded in the inner mitochondrial membrane, the thylakoid membrane of chloroplasts, and in the plasma membrane of bacteria. The membrane-bound $F_o$ part of ATP synthase contains the proton translocation pathway, and the $F_1$ part accommodates three catalytic binding sites for ATP or ADP plus phosphate, respectively, at the interface of the β and α subunits (Fig. 1).

Since the concept of the 'binding change mechanism' had been suggested by P. Boyer[1] subunit rotation of the two distinct motors of $F_o F_1$-ATP synthase is investigated with biochemical[2] and spectroscopical[3] methods. The rotary catalytic mechanism was supported by structural details of the $F_1$ part of ATP synthase obtained by X-ray crystallography[4], and in 1997 direct videomicroscopic evidence of ATP-driven rotation was provided using surface-immobilized single $F_1$ parts[5]. The proposed application of the $F_1$ part as a biological rotary nanomotor in bionanotechnology[6] was followed by the unraveling of substeps using nanometer sized beads as a marker of rotation[7].

The rotational mode of the $F_o$ motor has been studied by confocal single-fluorophore anisotropy[8] and by videomicroscopy of single beads attached to $F_o$[9, 10]. We apply a single-molecule FRET approach[11-13] using two fluorophores to monitor the rotary motions of the $F_o$ motor. The FRET donor is covalently attached to the non-rotating $a$ subunit of the $F_o$ motor by genetic fusion of the autofluorescent protein 'enhanced GFP' to the C terminus of $a$ [14]. Rhodamine derivatives are used as the FRET acceptor. 'Alexa-568' (or 'Atto-594', respectively) is bound to a single cysteine in the rotating ring of 10 $c$ subunits[15]. Rotation of the $c$ ring will result in a sequence of distance changes between these two fluorophores, and, accordingly, the FRET efficiency is expected to change stepwise (Fig. 1). The determination of FRET efficiency levels and the dwell times were done manually in the time trajectories of double-labeled ATP synthases, which were reconstituted into liposomes and were freely diffusing through the confocal single-molecule detection volume.

*m.boersch@physik.uni-stuttgart.de; phone (49) 711 6856 4632; fax (49) 711 6856 5281; www.m-boersch.org

Our preliminary single-molecule FRET data indicate a stepwise movement of the *c* ring during ATP-hydrolysis driven rotation that likely corresponds to three-step rotation with a stepsize of about 120° (Fig. 2a,b) in agreement to the single-molecule anisotropy data[8] and the videomicroscopy of single beads[10]. In contrast, proton-driven *c* ring rotation during ATP synthesis seems to involve smaller stepsizes, i.e. 10 substeps per full rotation with a stepsize of 36° (Fig. 2c,d). However, during the same measurements also larger stepsizes were detected that could correspond to 108° or 144° steps. To overcome the limitations of a potentially biased manual single-molecule FRET data analysis, a robust software approach is needed.

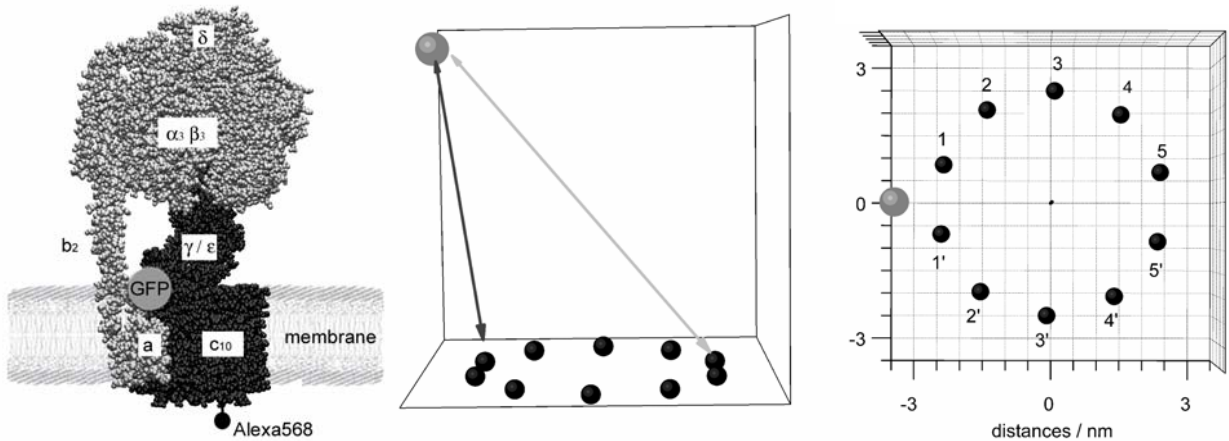

Fig. 1. Models of the FRET-labeled $F_oF_1$-ATP synthase embedded in a lipid vesicle and of the FRET distance changes upon *c* ring rotation. Left, the $F_1$ motor consists of the non-rotating subunits $\alpha_3\beta_3\delta$ and the rotating $\gamma$ and $\varepsilon$ subunit. The $F_o$ motor comprise the static subunits $ab_2$ and the rotating ring of 10 *c* subunits. The FRET donor EGFP (gray dot) is directly bound to subunit *a*. Only one *c* subunit is labeled with the FRET acceptor 'Alexa-568' (or 'Atto-594') on the opposite side of the membrane (black dot). Middle, the FRET efficiency between a nearest position of the labeled *c* and EGFP is high (black arrow) and low for the most distant position of c (light gray arrow). Right, evenly spaced arrangement of the ten possible positions of the FRET acceptor at one *c* subunit (bottom view). During *c* ring rotation FRET efficiency changes according to 36° substeps (1→2→3→4→) can be estimated and discriminated from larger stepsizes (1→5→3'→1→).

## 2. EXPERIMENTAL PROCEDURES

### 2.1 FRET-labeled $F_oF_1$-ATP synthase

Preparation procedures of the FRET-labeled $F_oF_1$-ATP synthase from *Escherichia coli* to monitor *c* ring rotation will be described in detail elsewhere. Briefly, we used a genetic fusion of the 'enhanced Green Fluorescent Protein', EGFP, to the C terminus of subunit *a* that has been constructed by Y. Bi (University of Western Ontario, Canada), combined with the additonal cysteine mutation at the $c_2$ position in the *c* subunit constructed by D. J. Cipriano (University of Western Ontario, Canada). $F_oF_1$-ATP synthase was purified from bacterial membranes as described[11-13]. Attaching the Alexa-568 (Molecular Probes) or Atto-594 (Atto-tec) to a single cysteine of detergent-solubilized $F_oF_1$ was accomplished a 4° C using the maleimide derivatives of the fluorophores. Rhodamine labeling efficiency was determined by UV-VIS absorption spectroscopy using the EGFP absorbance as an internal concentration reference. After labeling, ATP hydrolysis rates were measured as a control. Enzymes were reconstituted in an excess of preformed liposomes (radii of about 100 nm) to assure the ratio of a single enzyme per liposome. The FRET-labeled $F_oF_1$-ATP synthases in liposomes were stored as 5 µl aliquots at -80° C.

## 2.2 Confocal single-molecule FRET measurements

Single-molecule FRET measurements were performed on a confocal inverted microscope in epi-fluorescence configuration (Olympus IX71) using an argon ion laser for cw excitation at 488 nm (model 2020, Spectra Physics) or an amplified frequency-doubled laser diode for pulsed excitation at 488 nm (~ 60 ps pulses at 80 MHz, PicoTA 490, Picoquant)[14]. A water immersion objective (UPlanSApo 60x, 1.2 N.A., Olympus) focused the attenuated laser beams into a droplet of buffer containing 10 to 100 pM $F_oF_1$. Fluorescence from single $F_oF_1$ diffusing through the excitation volume of about 5 fl was split into two or three spectral ranges by a combination of consecutive dichroic mirrors (AHF) and detected by three avalanche photodiodes (SPCM AQR-14, Perkin Elmer), APDs, after passing interference filters (AHF). Out-of-focus fluorescence was rejected by a 150 μm pinhole. To investigate the photophysics of the FRET acceptor Alexa-568, confocal fluorescence correlation spectroscopy, FCS, was performed independently using cw excitation at 561 nm (solid-state laser Jive, Cobolt).

The signals of the APDs were counted by TCSPC (time correlated single photon counting) electronics from Becker & Hickl. For the two-APD configuration of single-molecule FRET measurements with high photon count rates, we installed the novel SPC-152 device consisting of two synchronized PC cards. FCS measurements were feasable with count rates up to 2 MHz per channel. A polarizing beam splitter was integrated into the detection pathway in order to measure the cross correlation functions for correlation times from 10 ns to 1 s in arbitrary parts of the time trajectories using the synchronized 'macro time' information of each photon. For three and four channel detection, we employ a SPC-630 PC card with an 8 channel router as described previously.

FRET data analysis was performed after binning the photons of the multichannel time trajectories to 1 ms intervals by our software 'Burst_Analyzer'[16]. FRET distances were determined from the proximity factor trajectory, that is, the ratio of the corrected fluorescence intensity of the acceptor channel, $I_A$, and the sum of donor and acceptor intensities, $I_D+I_A$, binned in 1 ms intervals. Alternatively, a maximum likelihood estimation approach based on photon distribution statistics was used yielding the FRET distance[17]. FCS curves were calculated using a fast multiple tau algorithm[18].

## 2.3 Simulation of confocal single-molecule FRET data

Single-molecule fluorescence data of a freely diffusing single particle, that is, a single fluorophore or the FRET-labeled $F_oF_1$ in a spherical liposome with a radius of 100 nm, was simulated by a 'random walk' through a virtual laser focus (or detection volume) in a three dimensional box. The time interval for the Brownian motion simulation was set to 100 μs. The detection volume was defined by three dimensional Gaussians and placed within the center of the box. Dimensions of the confocal detection volume were taken from the experiment resulting in a prolate ellipsoid with a lateral radius of 650 nm (intensity decayed to $1/e^2$ in the x-y dimensions) and an axial radius of 3.3 μm (intensity decayed to $1/e^2$ in the z dimension). The volume of the ellipsoid was 5.84 fl using the $2\sigma$ values which corresponds to 13.5 % of the intensity maximum. The diffusion of the single particle was restricted to a box with a length of 1.95 μm in x and y direction, and a height of 9.90 μm in z corresponding to a total volume of 37.64 fl. The diffusion properties of the particle were chosen according to the experimental data, that is, the diffusion constant of the $F_oF_1$ in a spherical liposome was 100 times smaller than the diffusion constant of a single fluorophore. When the particle hit a wall of the box, it was forced to re-enter the box in the next simulated time frame from the opposite side. The maximum fluorescence intensity of a particle in the center of the detection volume was set to 100 counts/ms. The concentration of the particles was kept in the picomolar range resulting in a single particle in the box at any time. 10 time trajectories of 10 s duration each were simulated and visualized using the software 'Burst_Analyzer'. The intensity autocorrelation functions of these simulated trajectories were calculated, and the mean number of molecules in the detection volume as well as the diffusion constants of the particle were calculated and compared to the initial values of the Monte Carlo simulation. This simulation was programmed in MATLAB (Mathworks, Inc.).

Fluctuating FRET efficiencies within the bursts were generated by a superposition of the intensity trajectory with the FRET level trajectory of a single particle. The FRET level trajectory was calculated using a sequence of three (or five) FRET levels. The dwell time of a FRET level and the transition probabilities between adjacent FRET levels were given by the ratio of rate constants for forward and backward transitions using experimental data from single-molecule FRET measurements of proton-driven rotation in $F_oF_1$. The simulated photon bursts with FRET level transitions were passed to the software 'Burst_Analyzer' and bursts were defined using intensity based thresholds. FRET level analysis including dwell time calculation for the intermediary FRET levels were accomplished by procedures published previously[14] for comparison with the HMM analysis.

**2.4 Hidden Markov Models and FRET data analysis**

Marked bursts containing proximity factor trajectories with 1 ms time resolution were the output from the 'Burst_Analyzer' and were further analyzed using Hidden Markov Models. The HMM analysis software was written in MATLAB and is based on the toolbox provided by K. Murphy[19]. Given the number of HMM states representing the FRET levels in the simulated bursts, the software optimized the mean values of the FRET levels from starting values and provided the dwell times as well as the transition probabilities. For the three FRET level simulation and for the five FRET level simulation data, a series of HMMs with different numbers of one to eight hidden states were used and compared. The most likely number of hidden states, that describe the simulated data set best, is found by the Bayesian information criterion, BIC.

# 3. CONFOCAL SINGLE-MOLECULE FRET MEASUREMENTS OF FREELY DIFFUSING $F_OF_1$-ATP SYNTHASE

Unraveling the stepsize of linear and rotary biological motors with nm precision requires long observation times for the individual stopping position of the single motor as well as long time trajectories that allow for observation of multiple turnover steps for appropriate statistics. This has been achieved for the ATP-driven $F_1$ motor of ATP synthase using a surface-attached $F_1$ and a polystyrene bead (or gold bead) as a marker of rotation, which was not photobleached during the measurement time of more than one hour. The same method was applied to the $F_o$ motor during ATP-driven rotation of the $c$ ring.

**3.1 $c$ ring rotation in $F_oF_1$-ATP synthase monitored by single-molecule FRET**

However, $F_oF_1$-ATP synthase has to be embedded into a lipid membrane and a pH difference has to be generated to monitor proton-driven rotation of the $F_o$ motor during ATP synthesis. Rotation of the $c$ ring was successfully monitored using the fluorescence anisotropy of a single Cy3 attached to one $c$ subunit of a $Na^+$-driven ATP synthase. The low photon count rate limited the resolution of the stepsize in this approach.

We use single-molecule FRET to monitor $c$ ring rotation in $F_oF_1$-ATP synthase during catalysis (Fig. 2). As the FRET donor, EGFP was permanently fused to the C terminus of subunit $a$ in the $F_o$ part. EGFP is a protein with a molecular mass of 27 kDa, but the fusion to the C terminus of subunit $a$ did not block the rotary movements of the $\gamma$ and $\epsilon$ subunits nor the $c$ ring. The FRET acceptor was either 'Alexa-568' or 'Atto-594'. These rhodamine derivatives were bound to a cysteine at residue position 2 in one $c$ subunit. The diameter of the $c$ ring is about 5 nm, and the expected height of the EGFP chromophore is approximately 4.2 nm above the level of rotation of the FRET acceptor. Accordingly, distance changes between the two fluorophores in $F_o$ are expected to occur from 4.2 nm up to 7.5 nm. This is the ideal distance change to be measured accurately by FRET assuming a Förster radius between 5 and 6 nm. The single-molecule FRET efficiency is calculated as the proximity factor P.

The P values were found to change between 0.1 and 0.9 during ATP hydrolysis as well as ATP synthesis. While large steps (comprising 3 or more substeps) were observed during ATP hydrolysis, the FRET efficiency changes during proton-driven $c$ ring rotation could not be described unequivocally by either large stepping or a persistent ten-stepped rotation in 36° substeps.

**3.2 Intensity fluctuations and FRET level transition determination**

Due to the Brownian movement through the laser focus, the total fluorescence intensity is spatially dependent and fluctuates. For transient low count rates, the apparent proximity factor approaches values around 0.5 because of a background fluorescence detected on both FRET donor and acceptor channels.

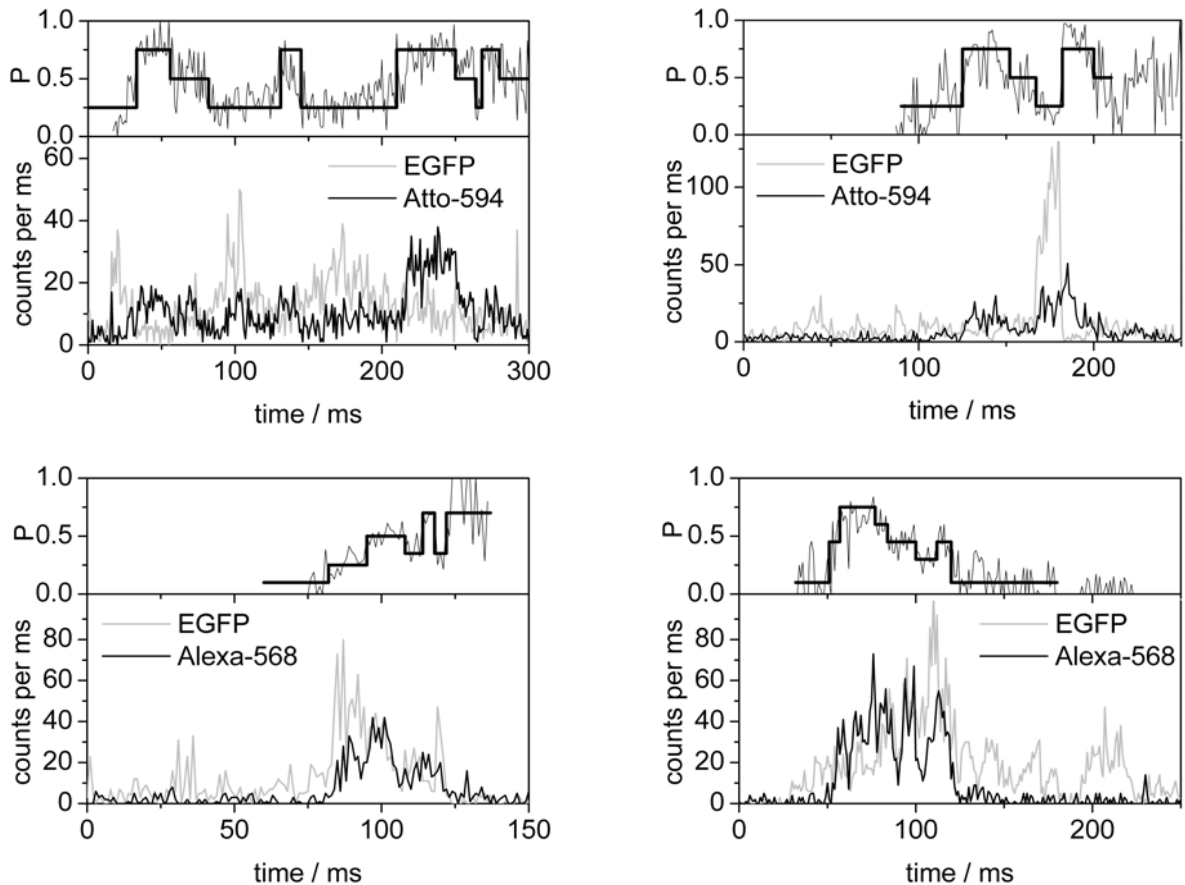

Fig. 2. Photon bursts and proximity factor trajectories of single FRET-labeled $F_oF_1$-ATP synthase in a liposome during catalysis. In the lower parts of the diagrams the gray traces display the FRET donor intensity binned in 1 ms, the black traces show the FRET acceptor intensity. Upper row, stepwise FRET efficiency changes during ATP hydrolysis (with FRET acceptor Atto-594). The large stepsize is defined by a repeating sequence of three FRET levels according to manual FRET data analysis. Lower row, FRET efficiency changes during ATP synthesis. Multiple FRET distance changes occur before the starting first FRET level of the burst is reached again according to manual FRET data analysis.

## 4. THE HMM APPROACH TO ANALYZE SINGLE-MOLECULE FRET

The Hidden Markov Models approach has to identify the number of FRET levels and the transitions between these levels. The method accounts for the intensity fluctuations within a single photon burst. To develop and evaluate the use of HMMs we started with a simulation of photon burst of freely diffusing particles with known FRET level transition rates. In the following step, HMMs with predefined numbers of hidden states were trained on the simulated data set. Finally, the best model with a minimum number of HMM states was determined using the Bayesian information criterion.

### 4.1 Monte Carlo simulation of confocal FRET trajectories

**4.1.1** To simulate the intensity distribution in the confocal laser excitation volume, a three dimensional Gaussian was chosen and placed in the center of the simulation box. Three dimensional Brownian motion is related to the diffusion coefficient by[20]

$$D = \frac{\langle x^2 \rangle}{6 t_{res}} \quad (1)$$

with $x$, mean distance after a time interval $t_{res}$ (time resolution of the simulation, 100 μs). The time resolution has to be high enough to accurately map the trace of the particle in and out of the detection volume. Therefore, the mean fluorescence intensity is given by the spatial detection efficiency of the constant intensity $I_0$:

$$I(x,y,z) = I_0 \exp\left(-2(x^2+y^2)/\omega_0\right) \exp\left(-2z^2/z_0\right) \quad (2)$$

with $\omega_0$, lateral radius, and $z_0$, axial radius of the excitation volume[21].

**4.1.2** The double-labeled $F_oF_1$ exhibits a FRET level $i$ with $i=1...Q$. Q is the number of distinct FRET levels for a full turn of the rotary motor of $F_oF_1$. Each FRET level is defined by a given value of the proximity factor, $P_i$. We simulated a three FRET level system (Q=3, $P_1$=0.3, $P_2$=0.5 and $P_3$=0.7) and a five FRET level system (Q=5, $P_1$=0.2, $P_2$=0.35, $P_3$=0.5, $P_4$=0.65, $P_5$=0.8). FRET level transitions were allowed for neighboring FRET levels only.

The probability $k_{ii}$ to find the $F_oF_1$ remaining in the same FRET level $i$ after one time step $t_{res}$=100 μs of the simulation was related to the dwell time, $t_{dwell}$

$$k_{ii} = \exp(-t_{res}/t_{dwell}) \quad (3)$$

To obtain a preferred direction of rotation, the transition probabilities to the neighboring FRET levels $i_+=i+1$ and $i_-=i-1$ were related by

$$k_{ii_+}/k_{ii_-} = 20 \quad (4)$$

Furthermore, all transition probabilities from state $i$ had to sum up to 1

$$k_{ii} + k_{ii_+} + k_{ii_-} = 1 \quad (5)$$

**4.1.3** The time dependent, fluctuating fluorescence intensity $I(t)$ of $F_oF_1$ in the actual FRET level $i$ at time $t$ is devided into a FRET donor intensity, $I_D$, and a FRET acceptor intensity, $I_A$, by the time dependent proximity factor $P(t)$

$$I_D(t) = (1-P(t))I(t) \quad (6)$$
$$I_A(t) = P(t)I(t) \quad (7)$$

**4.1.4** We consider the number of detected photons on both channels at time $t$ to be randomly distributed due to shot noise, which is described by the distribution of Poisson ($\zeta$, probability density function to detect $k$ photons)[22]

$$\zeta_{channel}(I_{channel}(t),k) = \frac{I_{channel}(t)^k}{k!} \exp(-I_{channel}(t)) \quad (8)$$

We generated Poisson distributed random variables with the time dependent $I_D$ and $I_A$ to obtain the single-molecule time trajectory. In the same way, we added a background noise of $I^{noise}$=3 counts/ms for each channel.

**4.1.5** To validate the simulated photon burst trajectories we calculated the fluorescence autocorrelation function $G(\tau)$ of the FRET donor channel. From the fit we obtained the diffusion time, $\tau_{Diff}$, of $F_oF_1$ through the detection volume[23]

$$G(\tau) = 1 + \frac{1}{N}\left(\frac{1}{1+\tau/\tau_{Diff}}\right)\left(\frac{1}{1+(\omega_0/z_0)^2 \tau/\tau_{Diff}}\right)^{1/2} \quad (9)$$

$\tau_{Diff}$ corresponded to the typical times of experimental FCS measurements of $F_oF_1$ in liposomes, that is, about 26 ms in the simulation. The initial diffusion constant, $D$, was regained by[23]

$$D = \frac{\omega_0^2}{4\tau_{Diff}} \quad (10)$$

**4.1.6** For the next step of the simulation the time trajectories were re-binned to 1 ms time resolution, which corresponds to the standard time resolution of the single-molecule FRET experiments. The photon bursts were identified by threshold values after background subtraction of 3 counts/ms on both FRET channels. A minimum level of 10 counts/ms on FRET donor or acceptor channel started the burst identification and 4 counts/ms marked the outer boundaries of a burst. A minimum burst length threshold was set to 5 ms, and within one bursts, the intensities were allowed to drop below the minimum level for up to 15 ms. Longer off-times resulted in a separate burst. Examples of simulated bursts are shown in Fig. 4a,b. After building the list of bursts, the retrieval process of the internal FRET level transitions was started.

### 4.2 The HMMs based on Beta probability density functions of the proximity factor of a three (and a five) FRET level system

Hidden Markov Models consist of four components:

- Q, number of hidden states $S_i$
- $k_{ij}$, transition probabilities between state $S_i$ to state $S_j$, with i,j=1...Q
- $\alpha$, observables
- $f(\alpha)$, emission probability function

The simulation (as well as the single-molecule FRET experiment) provides two intensity trajectories. However, the intramolecular distance between the two fluorophores is the only observable for the HMM. The proximity factor

$$P = \frac{I_A}{I_D + I_A} \qquad (11)$$

is used as a measure for the distance. Considering the intensities $I_D$ and $I_A$ as independent Poisson variables, the probability density function $f(P)$ of the proximity factor $P$ is distributed according to the Beta law[24, 25]

$$f(P) = \frac{1}{B(I_A, I_D)} P^{I_A-1}(1-P)^{I_D-1} \qquad (12)$$

with the Beta function $B$ defined by Gamma functions

$$B(I_A, I_D) = \frac{\Gamma(I_A)\Gamma(I_D)}{\Gamma(I_A + I_D)} \qquad (13)$$

The mean $\langle f(P) \rangle$ remains $\langle f(P) \rangle = \frac{I_A}{I_D + I_A}$. $\qquad (14)$

The standard deviation $\sigma$ of the Beta function is given by

$$\sigma = \left[ \frac{I_D I_A}{(I_D + I_A)^2 (I_D + I_A + 1)} \right]^{1/2} \qquad (15)$$

The presence of a constant background noise $I_i^{noise}$ on both channels shifts the expectation value of the proximity factor

$$\langle f(P) \rangle = \frac{(I_A + I_A^{noise})}{(I_D + I_D^{noise}) + (I_A + I_A^{noise})} \qquad (16)$$

Only for the case of high photon count rates, the distributions of the proximity factor from the experiment can be approximated by Gaussians. Then, the mean value, $\mu$, and the standard deviation, $\sigma$, describing the Gaussian can be expressed by the proximity factor $P$ and the sum of the intensities $I=I_D+I_A$ of both channels[24]

$$\mu = P \qquad (17)$$

$$\sigma = \frac{1}{\sqrt{I}} \left( P\sqrt{1-P} + (1-P)\sqrt{P} \right) \qquad (18)$$

Recently this approximation was used with an HMM applied to single-molecule FRET data[26]. However, this is only true, if the sum intensity $I$ is constant for each observable $\alpha$. The experimental confocal single-molecule FRET data of $F_oF_1$ in liposomes show strong fluctuations of the sum intensity due to Brownian motion in the detection volume. The different intensities result in different standard deviations. Furthermore, the approximation of the Beta function by a Gaussian is not sufficient for low count rates (Fig. 3).

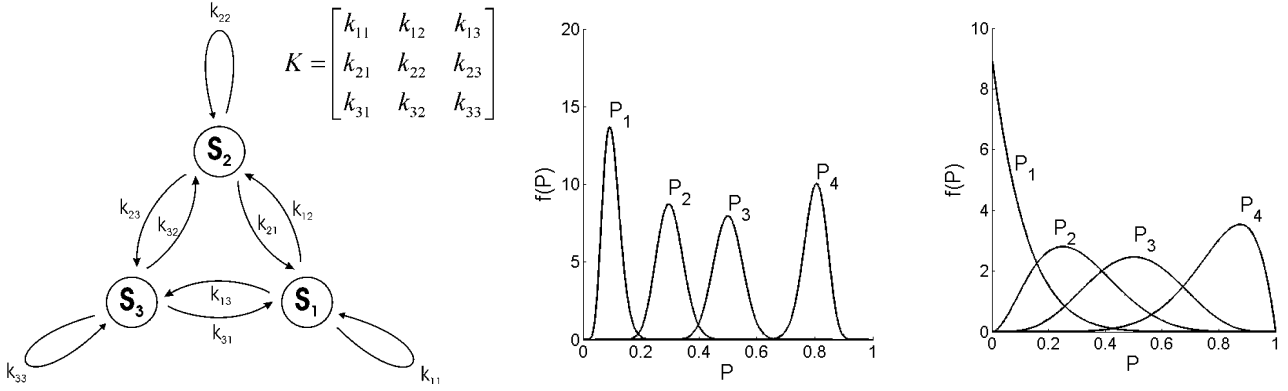

Fig. 3. Left, the Markov Model with three hidden states $S_i$ is described by the transition probability matrix $K$. Middle, beta probability density functions for proximity factors $P_1$=0.1, $P_2$=0.3, $P_3$=0.5, $P_4$=0.8 and a total intensity $I=I_D+I_A$=100 a.u.. The probability function could also be described by a Gaussian. Right, beta probability density functions for proximity factors $P_1$=0.1, $P_2$=0.3, $P_3$=0.5, $P_4$=0.8 and a total intensity $I=I_D+I_A$=10 a.u.. The shape of the probability functions $p_1$ and $p_4$ differ significantly from a Gaussian.

In order to describe the distribution of an observed variable $P$, the probability density function is built by a sum of N Gaussians described by the mean values $\mu_i$ and their standard deviation $\sigma_i$. The likelihood of the measured proximity factor of a data point can be expressed as

$$f(P|\mu,\sigma) = \frac{1}{(2\pi)^{N/2}|\sigma|^{1/2}} \exp(-d(P|\mu,\sigma)) \quad (19)$$

with the Mahalanobis distance $d(P|\mu,\sigma)$.

We replaced the multivariate Gaussian probability density function (eq. 19) by the Beta probability density function to characterize the distribution function of the measured proximity factor trajectory. In contrast to the Gaussian approximation, the shape of the emission probability function $f(\alpha)$ is adjusted to the actual intensity in every time step

$$f(P,I|\alpha) = \frac{1}{B(I_A,I_D)} P^{I_A-1}(1-P)^{I_D-1} \text{ with } I_D = (1-P)I \text{ and } I_A = PI. \quad (20)$$

The probability for a specific sequence of states $S$ is composed of the product of the emission probability $f$ at time $t$ and the transition probability $k$ from the state $S(t)$ to the next state $S(t+1)$:

$$L = \prod_t f(P(t),I(t)|\alpha) \, k_{S(t)S(t+1)} \quad (21)$$

With this equation it is possible to determine the likelihood $L$ of a given state sequence in a time trajectory of length $T$. The most likely state sequence given the data, the transition probability matrix and the emission probability functions, can be found within an order of $T$ as described[27]. Furthermore, the HMM parameters ($\alpha$, K) can be optimized by a maximum likelihood approach using the iterative Baum-Welch-algorithm[28]. Once the optimal parameters $\alpha$ and $K$ are estimated, the calculation of the Viterbi-path of all bursts yields the classification of each time step to a hidden state $S$. This procedure yields a statistical description of the 'hidden steps' of the rotational subunit movements within the ATP synthase.

By comparing several HMMs with a different number of hidden states, it is possible to decide, which model describes the data best, by comparing the overall likelihood of all bursts. This can be done by the Bayesian information criterion (BIC). The increased likelihood of a system with additional hidden states is compensated by a penalty term[29]:

$$BIC = -2\log L + (Q^2+1)\log \sum_{iBurst} A_{iBurst} \quad (22)$$

with $Q$, the number of hidden states (penalty term for redundant states) and $A_{iBurst}$, number of the assigned states within a single burst (penalty term for shortened dwell times).

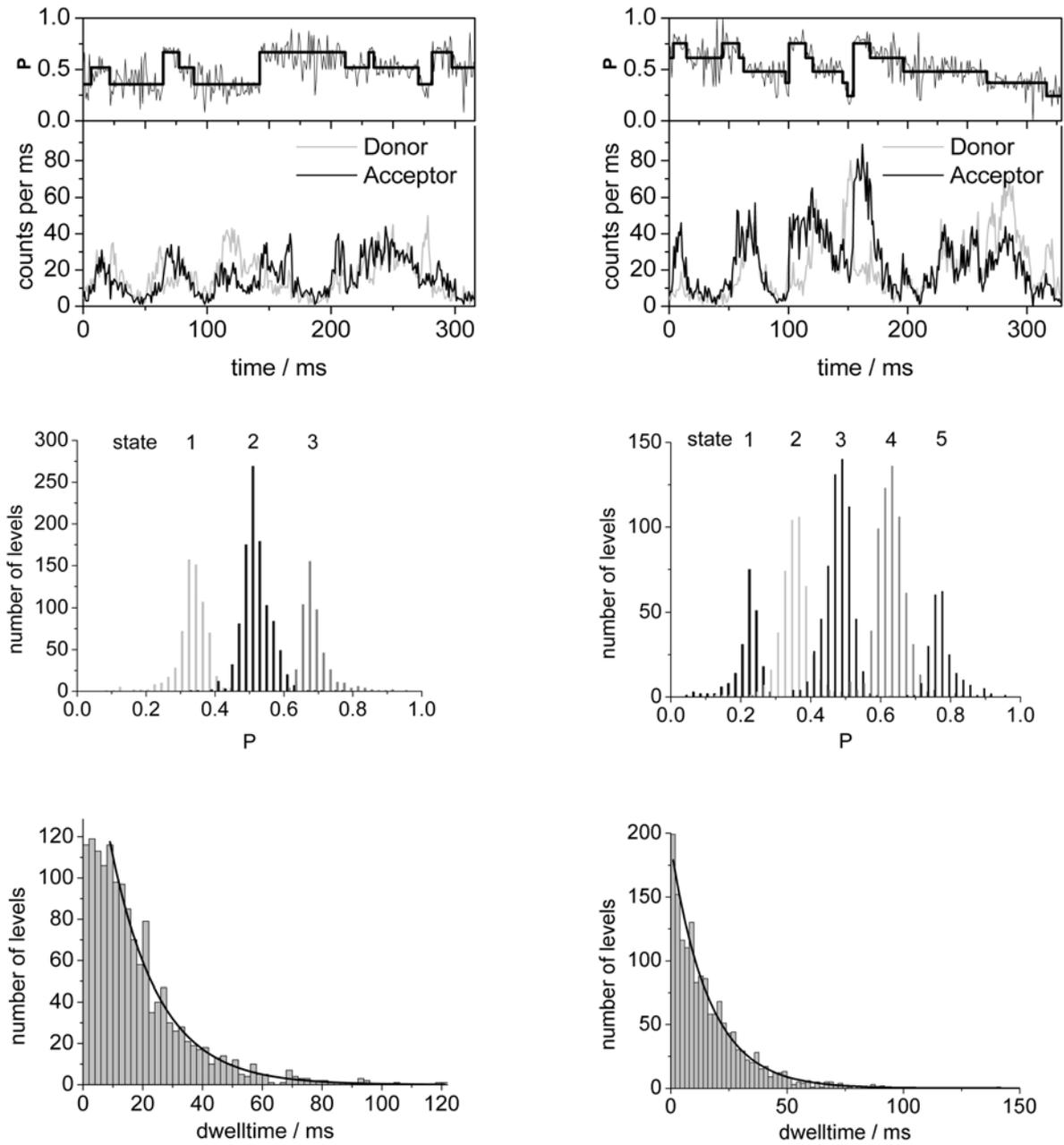

Fig. 4. Results of the HMM analysis with three hidden states $S_i$ of the simulated three FRET level system (left) and of the HMM with five hidden states $S_i$ of the simulated five FRET level system (right). Upper row, photon bursts of single simulated $F_oF_1$ with sum intensity fluctuations. The HMM extracted FRET levels of the complete data set are superimposed on the proximity factor trajectory of the single $F_oF_1$. Second row, FRET level histograms. Lower row, dwell time distributions of the intermediary FRET levels as found by the HMMs.

# 5. RESULTS

The results of the recovery processes for a given three and a five FRET level system by the HMM approach are shown in Fig. 4 and summarized in Table 1. The hidden state trajectory of the proximity factor had been recovered from the binned simulated data in the presence of Poisson distributed noise. The hidden states of a five level system had been resolved. Due to additional background noise, the resulting proximity factor of the hidden states corresponding to low and high proximity factors were shifted towards the value 0.5. This effect increases with decreasing signal-to-noise ratio of the simulated data.

In addition, background noise leads to a deviation of the proximity factor histogram (Fig. 4c,d) from the expected distributions (Fig. 3). Some parts of the burst have a count rate which is lower than the average count rate of the background noise. Due to the similar noise levels of both channels, these parts are best described by an HMM state with a proximity factor around 0.5. Therefore, the number of retrieved FRET levels with a proximity factory around 0.5 is higher than expected.

Table 1. HMM retrieval of proximity factors P and dwell times for the simulated FRET level system

| three FRET level system | | five FRET level system | |
|---|---|---|---|
| P (given for simulation): | 0.30, 0.50, 0.70 | P (given for simulation): | 0.20, 0.35, 0.50, 0.65, 0.80 |
| P (found by HMM with 3 states): | 0.36, 0.52, 0.67 | P (HMM with 5 states): | 0.24, 0.37, 0.48, 0.61, 0.75 |
| dwell time (given): | 20.0 ms | dwell time (given): | 20.0 ms |
| dwell time (found): | $16.1 \pm 0.6$ ms | dwell time (found): | $16.4 \pm 0.5$ ms |

The mean dwell times retrieved for the three and the five level systems were shorter than the initial values for the simulation. In principle the dwell times of the first and the last FRET level of a burst entering or leaving the detection volume remain unknown. Defining a burst by intensity thresholds leads to an additional shortening of the state trajectories due to splitting up into separate bursts. These effects shorten the average dwell time of the retrieved states and are not compensated in the current HMM approach.

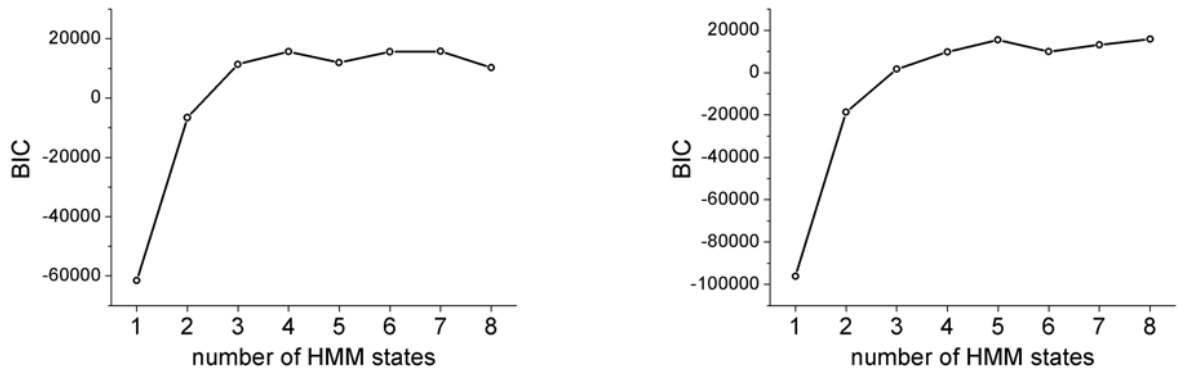

Fig. 5. The BIC for the three FRET level system (left) and the five FRET level system (right)

To determine the most likely HMM for the given FRET level system, the BIC (eq. 23) of different HMMs was applied for the three and five FRET level system (Fig. 5). The maximum for the three FRET level system was found for the HMM with 4 states, the maximum for the five FRET level system was found for the HMM with 5 states. The details of the HMMs are summarized in Table 2. We noticed that each supernumerous state leads to strong decrease in the corresponding dwell times. For example, the 4-state HMM for the three FRET level system found 4 states with P=0.33, 0.45, 0.55 and 0.67 with the corresponding dwell times 10.8, 5.5, 5.8 and 9.6 ms. The initial simulated state at P=0.50 was divided into the two states 0.45 and 0.55 with fast internal transitions. In general, more states resulted in very short dwell times (Tab. 2).

Table 2. Three and five FRET level systems with retrieved proximity factors and dwell times

| n-state HMM | three FRET level system with proximity factor P / dwell time (ms) | | | | | |
|---|---|---|---|---|---|---|
| 2 | 0.39 / 11.2 | 0.59 / 13.6 | | | | |
| 3 | 0.36 / 12.6 | 0.52 / 16.6 | 0.67 / 10.0 | | | |
| 4 | 0.33 / 10.8 | 0.45 / 5.5 | 0.55 / 5.8 | 0.67 / 9.6 | | |
| 5 | 0.01 / 0.04 | 0.36 / 12.9 | 0.51 / 17.0 | 0.65 / 12.9 | 0.52 / 0.06 | |
| 6 | 0.41 / 0.03 | 0.33 / 10.4 | 0.45 / 5.6 | 0.54 / 6.8 | 0.67 / 8.8 | 0.51 / 0.04 |
| | five FRET level system with proximity factor P / dwell time (ms) | | | | | |
| 2 | 0.39 / 17.9 | 0.62 / 18.4 | | | | |
| 3 | 0.32 / 11.3 | 0.52 / 15.3 | 0.69 / 11.9 | | | |
| 4 | 0.27 / 7.7 | 0.43 / 12.3 | 0.58 / 12.3 | 0.73 / 8.0 | | |
| 5 | 0.24 / 6.3 | 0.37 / 10.8 | 0.48 / 15.5 | 0.61 / 12.3 | 0.75 / 8.1 | |
| 6 | 0.11 / 0.4 | 0.29 / 7.3 | 0.44 / 12.6 | 0.57 / 12.0 | 0.71 / 7.4 | 0.89 / 0.3 |

The next steps towards a software-based unbiased analysis of confocal single-molecule FRET data will include a fine tuning of the BIC to identify the initial number of FRET levels and to use experimental single-molecule FRET data of the rotary motors of $F_oF_1$-ATP synthase.

**Acknowledgements**

The authors thank D.J. Cipriano and Y. Bi (University of Western Ontario) for the genetics of the $F_oF_1$-ATP synthase mutants, P. Gräber and co-workers (University of Freiburg, Germany) for the help with enzyme preparations and reconstitution methods, W. Becker for the SPC-152 PC cards and D. von Gegerfelt (Cobolt) for the Jive laser. Financial support from the Deutsche Forschungsgemeinschaft (BO 1891/8-1 and BO 1891/10-1) and from the Landesstiftung Baden-Württemberg (Network of Competence: Functional Nanostructures) is gratefully acknowledged.